\documentclass[conference]{IEEEtran}
\IEEEoverridecommandlockouts
\usepackage{algorithm}
\usepackage{tikz}
\usetikzlibrary{fit,calc,positioning}
\usepackage{algpseudocode} 
\usepackage[braket]{qcircuit}
\usepackage{cite}
\usepackage{caption}
\usepackage{amsmath,amssymb,amsfonts}
\usepackage{braket}
\usepackage{graphicx}
\usepackage{booktabs}
\usepackage{subcaption}
\usepackage{textcomp}
\usepackage{makecell}
\usepackage{xcolor}
\usepackage{hyperref}
\usepackage{multirow}
\def\BibTeX{{\rm B\kern-.05em{\sc i\kern-.025em b}\kern-.08em
    T\kern-.1667em\lower.7ex\hbox{E}\kern-.125emX}}

\usepackage{amsmath,algorithm,tabularx}
\makeatletter
\newcommand{\multiline}[1]{%
  \begin{tabularx}{\dimexpr\linewidth-\ALG@thistlm}[t]{@{}X@{}} 
    #1
  \end{tabularx}
}
\makeatother

\begin{document}

\title{A Deep Reinforcement Learning Method for Multi-objective Transmission Switching} 
\author{\IEEEauthorblockN{Ding Lin, Jianhui Wang }
\IEEEauthorblockA{\textit{Department of Electrical and Computer Engineering} \\
\textit{Southern Methodist University}\\
Dallas, TX 75205\\
\{dinglin; jianhui@smu.edu\}}
\and
\IEEEauthorblockN{Tianqiao Zhao,  Meng Yue}
\IEEEauthorblockA{\textit{Interdisciplinary Science Department
} \\
\textit{ Brookhaven National Lab}\\
Upton, NY 11973 \\
\{tzhao1; yuemeng@bnl.gov\}}

}

\maketitle

\begin{abstract}

Transmission switching is a well-established approach primarily applied to minimize operational costs through strategic network reconfiguration. However, exclusive focus on cost reduction can compromise system reliability. While multi-objective transmission switching can balance cost savings with reliability improvements, feasible solutions become exceedingly difficult to obtain as system scale grows, due to the inherent nonlinearity and high computational demands involved. This paper proposes a deep reinforcement learning~(DRL) method for multi-objective transmission switching. The method incorporates a dueling-based actor-critic framework to evaluate the relative impact of each line switching decision within the action space,  which improves decision quality and enhances both system reliability and cost efficiency.  Numerical studies on the IEEE 118-bus system verify the effectiveness and efficiency of the proposed approach compared to two benchmark DRL algorithms.

\end{abstract}


\section{Introduction}
\IEEEPARstart{T}{opology}  control, commonly referred to as transmission switching, is a proven approach for enhancing power system performance by selectively reconfiguring the network through the strategic opening or closing of transmission lines. This technique is widely implemented to strengthen system reliability by improving voltage profiles~\cite{shao2005corrective}, reducing the need for load shedding~\cite{khanabadi2012optimal}, and alleviating line congestion~\cite{granelli2006optimal}. From an economic standpoint, transmission switching also helps reduce generation costs and transmission losses by optimizing power flows~\cite{fisher2008optimal},~\cite{bacher1988loss}.

Despite these advantages, implementing transmission switching often requires balancing the trade-off between cost reduction and system reliability~\cite{ostrowski2014transmission}. Cost-driven decisions, such as opening lines to minimize losses, can reduce network redundancy and limit alternative pathways, which may compromise the system's resilience to disturbances. Reference~\cite{zhang2013optimal} proposes a multi-objective optimization framework for transmission switching that  simultaneously minimizes generation costs and maximizes probabilistic reliability. Building on AC formulation, \cite{8320949} presents a multi-objective congestion management approach formulated as a mixed-integer nonlinear programming to minimize operating costs and enhance reliability. However, multi-objective transmission switching, especially with AC models, faces inherent challenges due to the nonlinearity and computational demands of solving high-dimensional, non-convex problems, which complicates the attainment of feasible solutions as the system scale increases. These limitations underscore the need for more advanced methods capable of navigating the complexities of multi-objective transmission switching.

Deep reinforcement learning (DRL) has shown promise within the power systems field for handling complex decision-making tasks. By enabling agents to autonomously learn optimal strategies through interaction with dynamic environments, DRL can adapt to the high-dimensional, nonlinear nature of power systems.  For example, ~\cite{kou2020safe} introduces a safe off-policy DRL algorithm to solve Volt-VAR control problems with operational constraints in distribution systems.   A Q-learning based DRL method is developed to determine  the optimal charging strategy for  electric vehicle charging scheduling~\cite{8521585}.  To balance power fluctuations and improve system reliability, \cite{lu2019incentive} presents a DRL algorithm for real-time, incentive-driven demand response. A DRL approach  is proposed in \cite{8534442} for load frequency control to address uncertainties from renewable energy sources and enhance frequency stability. While few DRL applications focus on transmission switching, these successful examples highlight the potential of DRL for tackling the multi-objective demands of transmission switching.

In this paper, a dueling network-based DRL approach is proposed to learn optimal line switching strategies for the transmission switching problem. This method seeks to  simultaneously  reduce operational costs, stabilize voltage levels, lower power losses, alleviate congestion, and decrease the need for load shedding while adhering to the constraints of AC power flow equations.

The contributions of this paper are summarized as follows:
\begin{itemize}
\item 
This work proposes a DRL algorithm within a  discrete dueling soft actor-critic~(DDSAC) framework to address  the multi-objective transmission switching problem. This approach enables more precise differentiation among actions by explicitly evaluating the relative impact of each line-switching decision.
\item  
The effectiveness and efficiency of the proposed method, in comparison to two benchmark DRL algorithms, are demonstrated through case studies on the IEEE 118-bus system.
\end{itemize}

\section{PROBLEM DESCRIPTIONS} \label{formulation}
The transmission switching problem involves determining the optimal configuration of a power transmission network  to enhance system performance by deciding which transmission lines to keep in service or switch out. This section formulates a transmission switching problem  with the objective of reducing generator operating costs, improving voltage stability, reducing transmission losses,  relieving line congestion, and mitigating load shedding simultaneously while satisfying the AC power flow equations. Therefore, the transmission problems is described as follows:
\begin{subequations}
\begin{align}
\min \quad & f =  w_1 \sum_{i \in \mathcal{G}} C_{G_i}\left(P_{G_i}\right) \label{cg}\\
&+ w_2 \sum_{i \in \mathcal{B}}\left[\max \left(0,\underline{V}-V_i\right)+\max \left(0, V_i-\overline{V}\right)\right] \label{cvv}\\
& +w_3 \sum_{(i, j) \in \mathcal{L}} \left(\frac{S_{i j}}{S_{i j}^{\text {max }}}-1\right)_{+} \label{cline}\\
& +w_4 \sum_{(i, j) \in \mathcal{L}}  P_{i j}^{\text {loss }} \label{closs} \\
& +w_5 \sum_{(i, j) \in \mathcal{L}=}\left(1-x_{i j}\right) \label{line} \\
\text{s.t.} \notag \\
&P_{G_i}-P_{L_i}=V_i \sum_{j \in \mathcal{B}} V_j\left(G_{i j} \cos \theta_{i j}+B_{i j} \sin \theta_{i j}\right) \label{g}\\
&Q_{G_i}-Q_{L_i}=V_i \sum_{j \in \mathcal{B}} V_j\left(G_{i j} \sin \theta_{i j}-B_{i j} \cos \theta_{i j}\right) \label{h},
\end{align}
\end{subequations}
where $P_{G_i}$ and $Q_{G_i}$ are the active and reactive power generation at bus $i$;  $P_{L_i}$ and $Q_{L_i}$ denote the active and reactive power load at bus $i$; $G_{ij}$ and $B_{ij}$ represent the conductance and susceptance between buses $i$ and $j$;  $V_i$ is the voltage magnitude at bus $i$, and  $\theta_{ij}$ is the voltage phase angle difference between buses $i$ and  $j$; $\mathcal{G}$, $\mathcal{B}$, and $\mathcal{L}$ denote the set of generators, buses, and transmission lines, respectively.

The objective function $f$ consists of six terms~\eqref{cg}-\eqref{line}, each  weighted by $w_k$. Eq.~\eqref{cg} denotes the operating cost of generators, where $C_{G_i}$ is the cost coefficient for generator $i$; Eq.~\eqref{cvv} captures the deviations in bus voltage magnitudes  outside the upper and lower limits, $\overline{V}$ and $\underline{V}$; Eq.~\eqref{cline} describes line loadings that exceed their thermal limits, where $S_{ij}$ and $S_{i j}^{\text {max }}$ are the power flow and  thermal limit on line $(i, j)$; Eq.~\eqref{closs} accounts for active power losses $P_{ij}^{\text{loss}}$ in transmission lines;  Eq.~\eqref{line} aims at reducing the number of lines to be switched off, where $x_{i j} \in\{0,1\}$ represents the switching status of line $(i, j)$;  Eqs.~\eqref{g}-\eqref{h} are the active and reactive power balance constraints.

To formulate the DRL-based transmission switching problem, we consider a Markov Decision Process~(MDP) $\mathcal{M}=(\mathcal{S}, \mathcal{A}, r , P, \gamma)$, which consists of a continuous state space $\mathcal{S}$, a discrete action space $\mathcal{A}$, a reward function $r:\mathcal{S} \times \mathcal{A} \to \mathbb{R}$, a transition probability function $P: \mathcal{S} \times \mathcal{A} \times \mathcal{S} \to \mathbb{R}^+$, and a discount factor $\gamma \in [0,1)$. In an MDP, an agent operates based on a policy function \( \pi: \mathcal{S} \to \mathcal{A} \), mapping the  state \( s \) to an action \( a \). The policy is trained using experiences collected through interactions with the environment over discrete time steps $t$. These time steps are grouped into episodes, which may have a finite or infinite number of steps. At each step \( t \), the agent observes the current state \( s_t \in \mathcal{S} \) of the environment and selects an action \( a_t := \pi(s_t) \in \mathcal{A} \). Following the agent’s action, the environment transitions to the next state \( s_{t+1} \) according to the transition probability \( P(s_{t+1} \mid s_t, a_t) \) and provides the agent with a reward \( r(s_t, a_t) \).

The objective of the agent is to find a policy $\pi$ that maximizes the expected return over a finite time horizon $T$: 
\begin{equation}
\max _\pi J(\pi) =  \underset{\substack{\left(s_t, a_t\right) \sim \tau_\pi}}{\mathbb{E}}\left[\sum_{t=0}^T\gamma^t \, r\left(s_t, \pi(s_t) \right)\right],
\end{equation}
where $\tau$ is a trajectory induced by policy $\pi$. 

The total discounted reward  from a given time step $t$ is defined as $ G_t = \sum_{k=t}^T \gamma^{k-t} r\left(s_k, a_k\right)$. The value function represents the expected discounted return under a given policy $\pi$ when starting from state $s$:
\begin{equation}
    V_\pi(s) = \mathbb{E}_\pi\left[\sum_{k=t}^T \gamma^{k-t} r\left(s_k, a_k\right) \mid s_t = s \right].
\end{equation}
Similarly, the action-value function gives the expected discounted return for taking action $a$ in state $s$ under policy $\pi$:
\begin{equation}
    Q_\pi(s, a) = \mathbb{E}_\pi\left[\sum_{k=t}^T \gamma^{k-t} r\left(s_k, a_k\right) \mid s_t = s, a_t = a \right].
\end{equation}

In this MDP, the environment is represented by the transmission network, and the agent~(i.e., topology controller) interacts with this environment to learn optimal topology control strategies. The state space includes the active power output $p_G$ of generators, voltage magnitudes $v$ at various buses, line flows $s$, active power losses $p_{\text{loss}}$, loads $p_L$, and line status $x$. Therefore, the state at time step $t$ is represented as:
\begin{equation}
s_t=\left[p_G^t, v^t, s^t, p_{\text{loss}}^t, p_L^t, x^t, t\right].
\end{equation}
The action space  consists of line switching decisions. Thus, the action at time step $t$ is defined as:
\begin{equation}
a_t = \left[x_1^t, x_2^t, \cdots, x_{N_L}^t\right],
\end{equation}
where $N_L$ is the number of transmission lines.
To penalize divergence in power flow and mitigate the risk of islanding, the reward function is defined as follows:
\begin{equation}
r(s_t, a_t)= \begin{cases}-R_e, & \text {if islanding or power flow diverges } \\ 
- f, & \text { otherwise }\end{cases}
\end{equation}
where $-R_e$ is a large penalty applied in cases of islanding or power flow divergence. Otherwise, the reward is calculated as the negative of the objective function defined in Eqs.~\eqref{cg}-\eqref{line}. 

\section{ Discrete Dueling SAC algorithm for Transmission Switching} 
\label{algorithm}
\begin{figure}
 \centering
 \includegraphics[width=\linewidth, ]{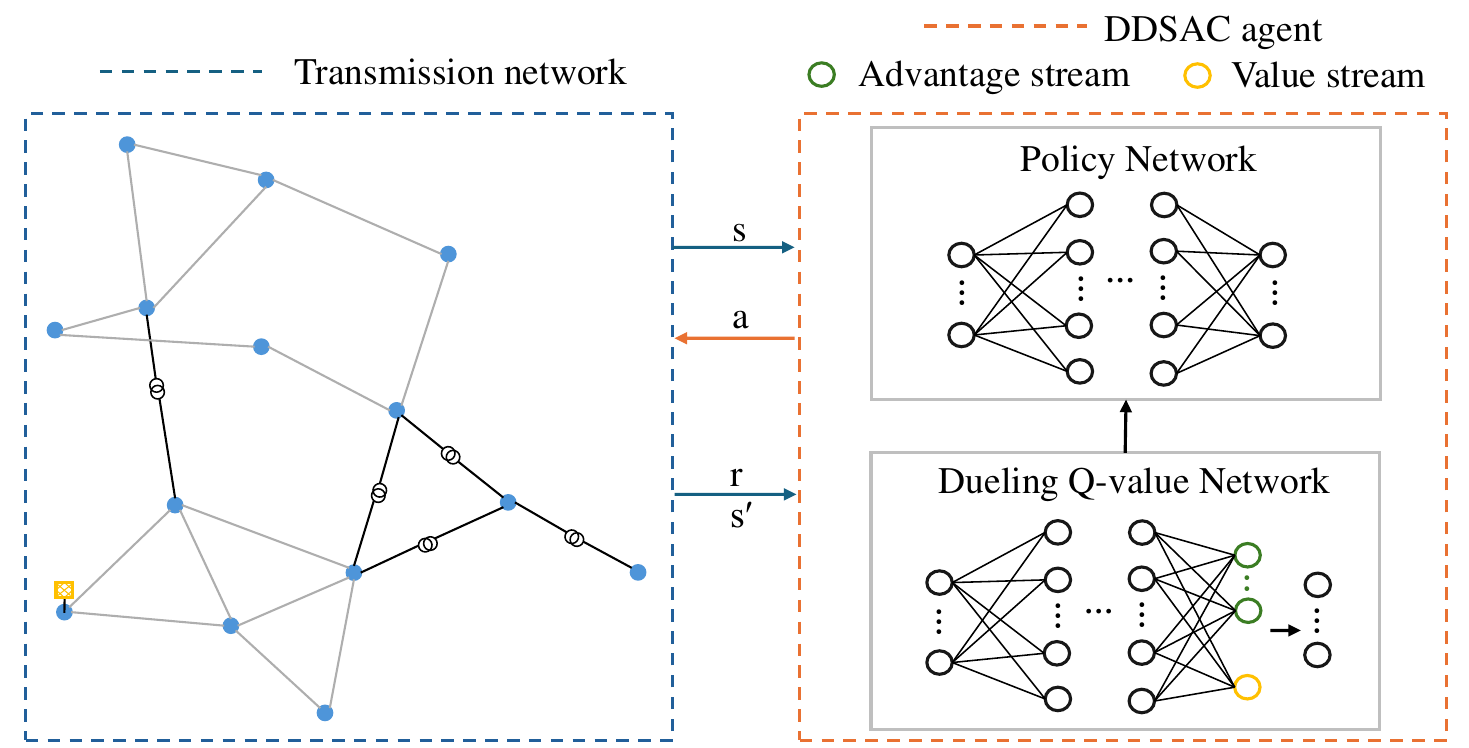}
 \caption{Proposed DDSAC framework. The environment (i.e., transmission network) interacts with the agent to learn optimal transmission switching strategies.}
\label{ots}
\end{figure}

This section introduces the proposed DDSAC algorithm for transmission switching problems, the framework of DDSAC is illustrated in Fig.~\ref{ots} . We begin with an overview of the classical SAC algorithm, followed by a detailed description of the  DDSAC algorithm.
\subsection{Soft Actor-Critic for Continuous Tasks}
SAC is a leading reinforcement learning algorithm developed for optimal decision-making in complex environments~\cite{haarnoja2018soft}. Operating within a maximum entropy framework, SAC encourages exploration by jointly maximizing the expected cumulative reward and the entropy of policy. This balance is achieved through soft policy iteration, which alternates between evaluating and refining the policy to improve performance and ensure stability:
\begin{equation}
\max _\pi J(\pi)= \underset{\substack{\left(s_t, a_t\right) \sim \tau_\pi}}{\mathbb{E}}\left[\sum_{t=0}^T\gamma^t\left(r\left(s_t, a_t\right)+\alpha \mathcal{H}\left(\pi\left(. \mid s_t\right)\right)\right)\right],
\end{equation}
where $\alpha$, known as the \emph{temperature} parameter, controls the importance of the entropy term relative to the reward and modulates the level of exploration; The policy $\pi$ is represented as a probability distribution to capture the inherent stochasticity required for entropy maximization; The entropy term $\mathcal{H} (\pi\left( \cdot \mid s_t\right))$ encourages exploration by rewarding  action diversity:
\[
\mathcal{H}(\pi\left( \cdot \mid s_t\right)) = \mathbb{E}_{a_t \sim \pi}\left[-\log \pi\left(a_t \mid s_t\right) \right]
\]

In practice, SAC approximates the \emph{optimal} action-value function \( Q^* \) through a neural network \( Q_\phi \), where \( Q_{\phi}(s, a) \approx Q^*(s, a) \) for each state-action pair \( (s, a) \), with \( \phi \) representing the parameters of the Q-network. The policy function is likewise approximated by a separate neural network \( \pi_\theta \), parameterized by \( \theta \), to closely approximate the optimal policy.




\subsection{DDSAC  for Transmission Switching Problems}

To address the discrete action space in the transmission switching problem, the SAC algorithm, originally designed for continuous actions, is adapted into the DDSAC framework. 

In DDSAC, the policy network generates a probability distribution over discrete actions to allow  probabilistic selection of line statuses. The Q-value network is split into two separate streams: a value stream and an advantage stream~\cite{wang2016dueling}. The value stream assesses the overall value of a specific state and captures the shared value of the state independent of the actions taken.  The advantage stream, on the other hand, estimates the relative advantage of each action in the current state and compares each action to the average performance. The final Q-value in DDSAC for each action is then computed as:
\begin{equation}
Q(s, a)=V(s)+\left(A(s, a)-\frac{1}{|\mathcal{A}|} \sum_{\tilde{a}} A\left(s, \tilde{a}\right)\right),
\end{equation}
where $V(s)$  is the value function for state $s$, $A(s, a)$ denotes advantage function for state $s$ and action $a$, and $\tilde{a}$ represents a possible action from action space $\mathcal{A}$. This formulation helps in better distinguishing among actions by explicitly identifying the relative impact of each action, which is useful for discrete line switching decisions in transmission networks.

DDSAC employs two Q-value networks \( Q_{\phi_1} \) and \( Q_{\phi_2} \), with parameters \( \phi_1 \) and \( \phi_2 \), to reduce overestimation bias. A separate target network \( Q_{\text{target}} \) is also maintained for each Q-value network, with target parameters \( \phi_1' \) and \( \phi_2' \) updated periodically. The Q-value loss function aims to minimize the difference between the current Q-value and the target Q-value. The target for Q-value update is computed as:
\[
y(r, s', a') = r +  \gamma \pi_\theta(s', a')^\top[ \min_{i=1,2}  Q_{\phi_i'}(s', a')  - \alpha \log \pi(a' | s')]
\]
where \( s' \) is the next state,  \( a' \) is an action sampled from the current policy, and $\pi_\theta(s', a')$ represents the action probability distribution of policy.

The Q-value loss function is  the mean-squared error between the current Q-value $ Q_{\phi_i}(s, a)$ and the target $y(r, s', a')$:
\begin{equation}
\label{qvalue}
    \mathcal{L}_{Q_i}(\phi_i) = \mathbb{E}_{(s, a, r, s') \sim \mathcal{B}} \left[ \left( Q_{\phi_i}(s, a) - y(r, s', a') \right)^2 \right],
\end{equation}
where  \( \mathcal{B} \) is the replay buffer. After each training step, the target network parameters \( \phi_1' \) and \( \phi_2' \) are updated utilizing a soft update:
\begin{equation}
\label{target}
    \phi_i' \leftarrow \tau \phi_i + (1 - \tau) \phi_i',
\end{equation}
where \( \tau \) is a small constant  determining the update rate.

The policy network \( \pi_\theta \),  parameterized by \( \theta \), is trained to maximize the entropy-augmented return. This is achieved by minimizing the following objective to encourage high Q-values and policy entropy:
\begin{equation}
\label{policy}
    \mathcal{L}_\pi(\theta) = \mathbb{E}_{s \sim \mathcal{B}} \left[ \pi_\theta(s, a)^\top \left[ \alpha \log \pi_\theta(a | s) - \hat{Q}_{\phi}(s, a) \right] \right],
\end{equation}
where \( \hat{Q}_{\phi}(s, a) = \min \left( Q_{\phi_1}(s, a), Q_{\phi_2}(s, a) \right) \) is used to mitigate overestimation.

The temperature parameter \( \alpha \) is adjusted to maintain a target entropy \( \mathcal{H}_{\text{target}} \). The temperature loss is minimized to keep the policy's entropy close to \( \mathcal{H}_{\text{target}} \):
\begin{equation}
\label{alpha}
    \mathcal{L}_\alpha(\alpha) = \mathbb{E}_{s \sim \mathcal{B}} \left[ -\alpha \pi_\theta(s, a)^\top \left[ \log \pi_\theta(a | s) + \mathcal{H}_{\text{target}}\right] \right],
\end{equation}
where $\mathcal{H}_{\text{target}}$ is a hyperparameter representing the desired level of entropy:
\begin{equation}\label{hvalue}
\mathcal{H}_{\text{target}} = 0.25(-\log(\frac{1}{\rm{dim}(\mathcal{A})})).
\end{equation}

The overall framework of the proposed DDSAC algorithm is summarized in Algorithm~\ref{qsac}. 
\begin{algorithm}
\caption{DDSAC}
\label{qsac}
\begin{algorithmic}[1] 
\State \textbf{Input:} Initial $\theta$,  $\phi_1$ and $\phi_2$, and replay buffer  $\mathcal{B}$.
\State Initialize target Q-value network parameters: $\phi_1^{\prime} \leftarrow \phi_1$ and $\phi_2^{\prime} \leftarrow \phi_2$.
\State Set  target entropy $\mathcal{H}_{\text{target}}$ using Eq.~\eqref{hvalue}.
\For{each episode}
    \For{$t \leftarrow 1$ $\textbf{to}$ $T$}
        \State \multiline{Observe the state $s_t$,  and sample an action $a_t \sim \pi_\theta(\cdot \mid s_t)$.}
        \State \multiline{Observe the next state $s_{t+1}$, collect the reward $r_t$, and receive the terminal signal $d$ to determine if $s_{t+1}$ is terminal or not.}
        \State \multiline{Store sample $\mathcal{B} \leftarrow \mathcal{B} \cup\left(s_t, a_t, s_{t+1}, r_t, d\right)$.}
    \EndFor
    \For{each parameter updating step}
        \State Randomly sample a mini-batch from $\mathcal{B}$.
        \State \multiline{Update Q-value networks $Q_{\phi_i}$ by minimizing Eq.~(\ref{qvalue}).}
        \State \multiline{Update policy network $\pi_{\theta}$ by minimizing Eq.~(\ref{policy}).}
        \State Update temperature $\alpha$ by minimizing Eq.~(\ref{alpha}).
        \State Soft update target  networks $Q_{\phi^{\prime}_i}$ using Eq.~(\ref{target}). 
   \EndFor
\EndFor
\end{algorithmic}
\end{algorithm}

\begin{figure}[t]

    \centering
    \includegraphics[width=\linewidth]{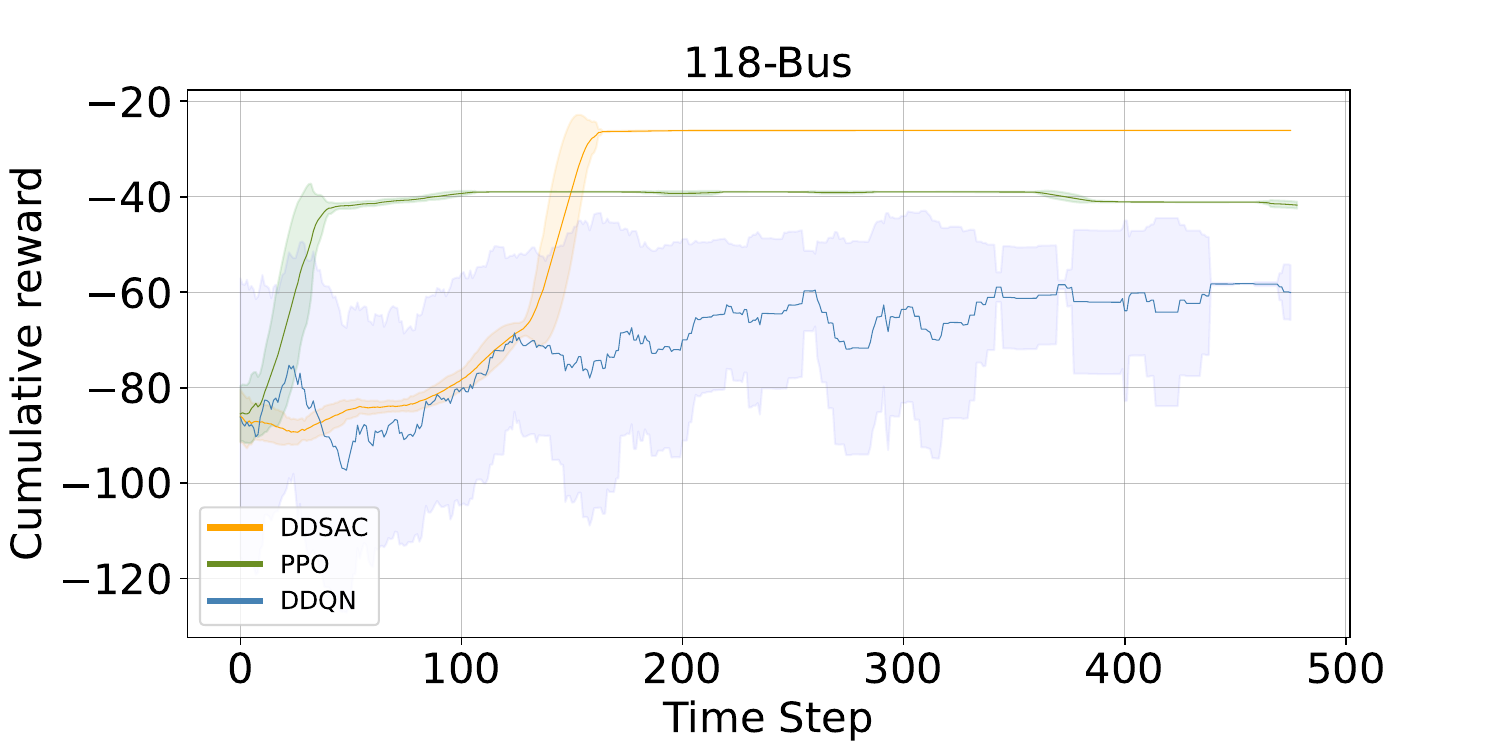}
    \caption{Cumulative reward in 118-bus system.}
    \label{118busreward}
\end{figure}
\section{Numerical Results}
\label{case}
In this section,   the effectiveness and efficiency of the proposed DDSAC algorithm is validated using the IEEE 118-bus system.  First, we set up simulations for two benchmark algorithms, Double DQN (DDQN) and Proximal Policy Optimization (PPO), alongside the proposed DDSAC. A comprehensive performance comparison between the two algorithms is then conducted using PyTorch.
\begin{figure*}[t]
    \centering
    \begin{subfigure}[b]{0.49\textwidth}  
    \centering
    \includegraphics[width=\linewidth]{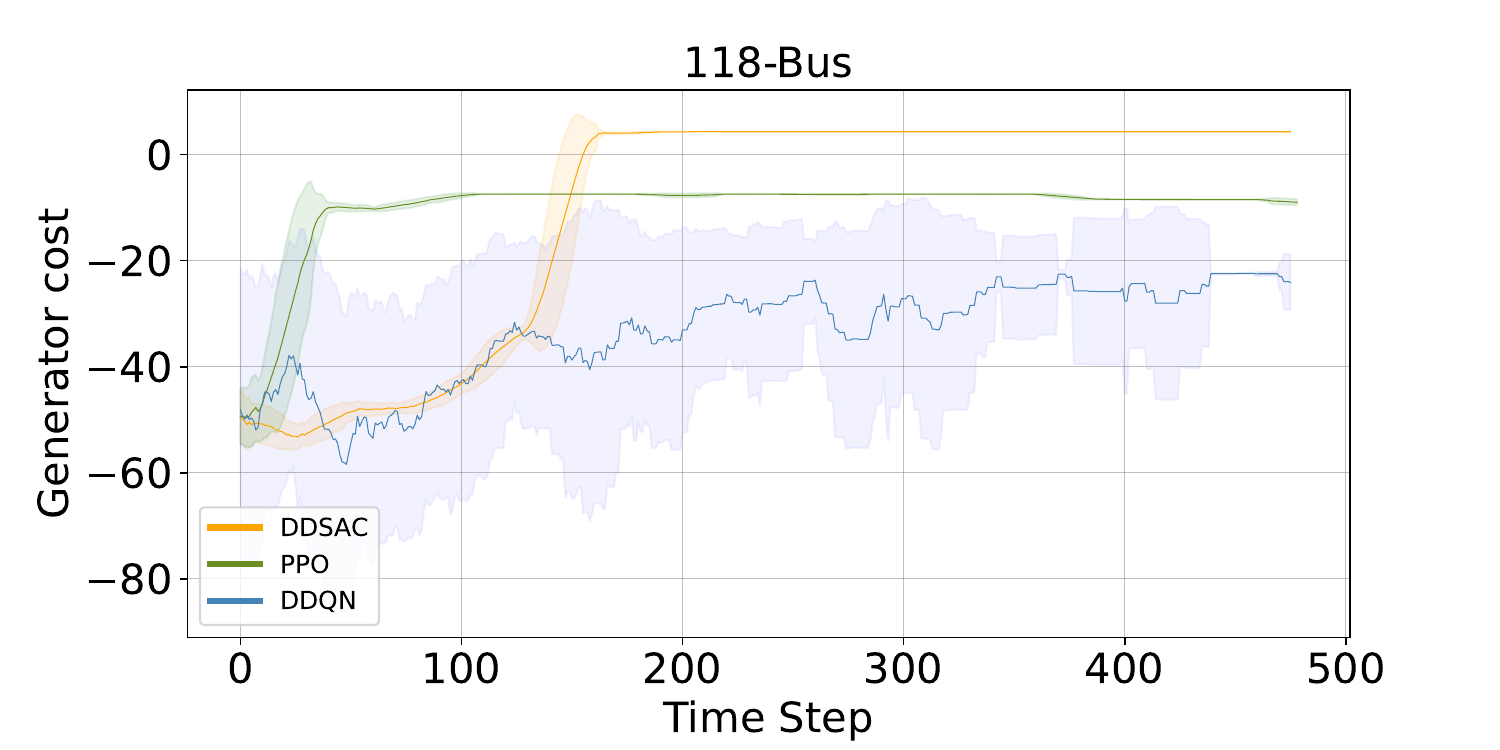}
    \end{subfigure}
    \hfill
    \begin{subfigure}[b]{0.49\textwidth}  
        \centering
        \includegraphics[width=\linewidth]{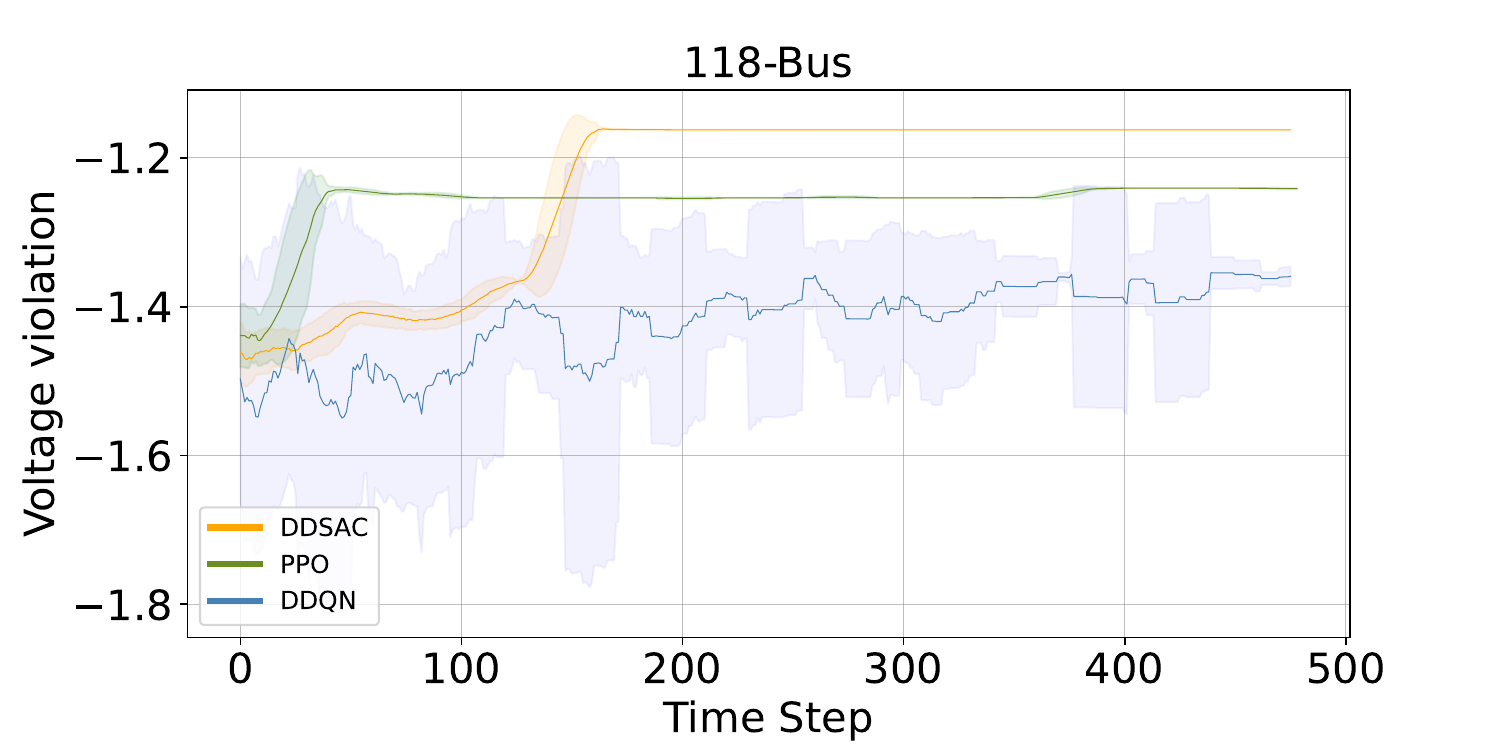}
    \end{subfigure}
    \hfill
    \begin{subfigure}[b]{0.49\textwidth}  
        \centering
        \includegraphics[width=\linewidth]{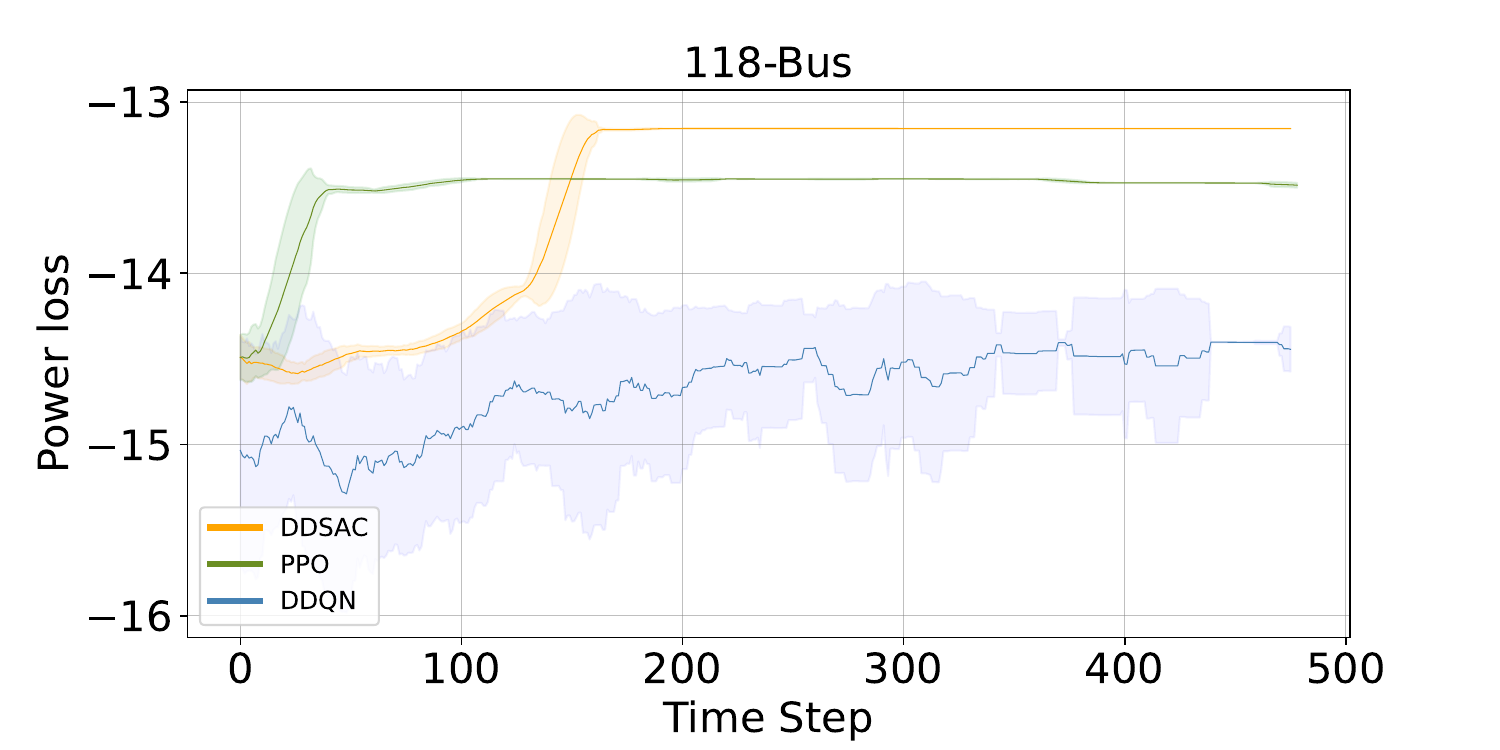}
    \end{subfigure}
    \hfill
    \vspace{0em} 
    \begin{subfigure}[b]{0.49\textwidth}  
        \centering
        \includegraphics[width=\linewidth]{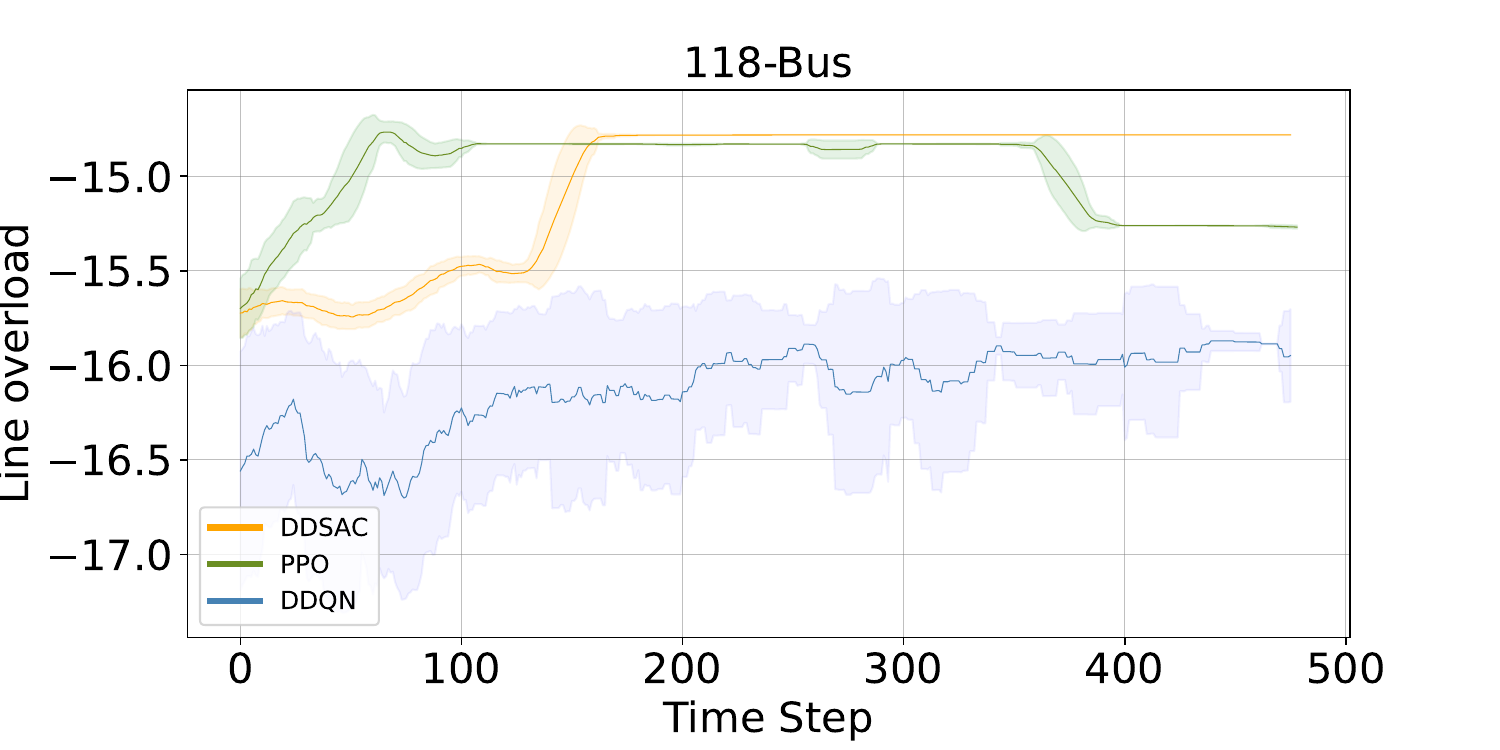}
    \end{subfigure}
    \caption{Generator cost, voltage violation, active power loss, and line overload  in 118-bus system.}
    \label{118bus}
\end{figure*}

\subsection{Simulation Setup}
To balance the weights of each reward component within the reward calculation, the generator cost term \eqref{cg} is calculated as the difference in total generator costs between the post-switching scenario and the initial scenario, where all lines are in service.  Voltage limits, $\overline{V}$ and $\underline{V}$, are set to 1.05 and 0.95 per unit.  Table~\ref{weight} outlines the values of each weight factor $w_1$ through $w_5$. 

To demonstrate the advantages of the DDSAC algorithm, we use DDQN and PPO  as  benchmarks. DDQN is an extension of the original DQN algorithm that helps mitigate the overestimation bias in Q-value updates by using separate networks to select and evaluate actions. PPO is a widely used policy-gradient method that optimizes policies with a clipped surrogate objective to maintain stability and improve sample efficiency. The hyperparameters for DDSAC, DDQN, and PPO are listed in Table~\ref{hyper}, with each algorithm’s hyperparameters individually tuned for optimal performance.

{
\renewcommand{\arraystretch}{1.1} 
\begin{table}[t]
    \centering
    \caption{Weight Factor for Reward Function}
    \label{weight}
    \scalebox{1}{
    \setlength{\tabcolsep}{12pt}
    \begin{tabular}{c|c}
        \hline
        \textbf{Weight Factor} & \textbf{Values} \\
        \hline
         $w_1$  & 0.1 \\
         $w_2$ & 100 \\
         $w_3$ & 1 \\
        $w_4$  & 0.1 \\
        $w_5$ & $10\cdot \frac{1}{N_L}$ \\[0.4em]
        \hline
    \end{tabular}
    }
\end{table}
}
{
\renewcommand{\arraystretch}{1.25} 
\begin{table}[t]
    \centering
    \caption{Hyperparameters of Three Algorithms}
    \label{hyper}
    \scalebox{1}{
    \setlength{\tabcolsep}{14pt}
    \begin{tabular}{l|l|c}
        \hline
        \textbf{Algorithms}& \textbf{Hyperparameters} & \textbf{Values} \\
        \hline
        \multirow{4}{*}{Shared Parameters} & number of hidden layers & 2 \\
        & number of hidden units  & 256 \\
        & learning rate  & 1e-4 \\
        & discount factor $\gamma$ & 0.99 \\
        \hline
        \multirow{3}{*}{DDSAC} 
        & $\mathcal{H}_{\text{target}}$ & Eq.~\eqref{hvalue} \\
        & soft update factor  $\tau$ & 0.005 \\
        & mini-batch size & 32 \\
        \hline
        \multirow{2 }{*}{PPO} 
        & clipping parameter & 0.2 \\
        & entropy coefficient & 0.02 \\
        \hline
        \multirow{3}{*}{DDQN}
        & mini-batch size & 32 \\
        & epsilon decay & 0.995 \\
        &target update frequency & 100 \\
        \hline
    \end{tabular}
    }
\end{table}
}

\subsection{Effectiveness and Efficiency}

To demonstrate the effectiveness and efficiency of the proposed algorithm, we compare the cumulative rewards, generator costs,  voltage violations, power loss, and line overload of the three algorithms, as shown in Fig.~\ref{118busreward} and Fig.~\ref{118bus}. Dark-colored curves represent the average performance over 10 runs, while the shaded areas indicate the standard error.

From the results in Fig.\ref{118busreward}, the DDSAC algorithm achieves faster convergence and a significantly higher cumulative reward compared to both PPO and DDQN. Within approximately 50 time steps, DDSAC stabilizes at a higher reward level, showcasing its superior learning efficiency in optimizing control actions for the 118-bus system. The stability of DDSAC is also evident, as its curve exhibits minimal fluctuations, unlike DDQN, which shows larger variations and a slower convergence rate.

In Fig.~\ref{118bus}, DDSAC consistently outperforms the other algorithms in minimizing generator costs, voltage violations, power losses, and line overloads, maintaining higher and more stable values throughout the time span. Its reduced fluctuations and superior performance across all metrics highlight DDSAC’s effectiveness, efficiency, and practicality for addressing multi-objective transmission switching problems.

\section{Conclusion}\label{Conclusions}

In this paper,   a dueling-based DRL algorithm is developed to efficiently solve the multi-objective transmission switching problem. By incorporating a dueling Q-value network within the discrete SAC framework, the algorithm provides a refined evaluation of line switching actions to enable more effective prioritization of critical decisions. Numerical studies on the IEEE 118-bus system demonstrate that the proposed algorithm achieves faster convergence and yields higher, more stable rewards than the benchmark DRL algorithms, which underscores the advantages of our approach.
\bibliography{article-template}

\end{document}